%% file: template_daga.tex
\pdfoutput=1

\documentclass[a4paper, 10pt, twocolumn]{article}
\def\correct<#1>{\textbf{* #1 *}}

\usepackage{graphicx}               % optional f�r Grafiken
\usepackage{tabularx}               % optional f�r Tabellen
\usepackage{multirow}               % optional f�r Tabellen
\usepackage{url}                % optional f�r Internet Links

\usepackage[small,bf]{caption2}     % bitte f�r Bildunterschriften verwenden

\usepackage{parskip}

\usepackage{titlesec}
\usepackage{amsmath}                % optional f�r Formeln

\titleformat{\section}{\normalfont\large\bfseries}{\thesection}{}{}
\titleformat{\subsection}{\normalfont\large\bfseries}{\thesection}{}{}
\titleformat{\paragraph}{\normalfont\bfseries}{\theparagraph}{}{}
\titlespacing{\section}{0pt}{6pt}{-1pt}
\titlespacing{\subsection}{0pt}{3pt}{-1pt}
\titlespacing{\paragraph}{0pt}{3pt}{-1pt}

\newcolumntype{Y}{>{\centering\arraybackslash}X}    %f�r Tabellen mit tabularx

\begin{document}

\date{}                                         % kein Datum auf 1. Seite

\title{\vspace{-8mm}\textbf{\large
Investigations on the Influence of Combined Inter-Aural Cue Distortions on Overall Audio Quality }}

% Hier die Namen und Daten der beteiligten Autoren eintragen
\author{
Pablo M. Delgado$^1$, Jürgen Herre$^{1,2}$\\
$^1$ \emph{\small International Audio Laboratories Erlangen, 90158 Erlangen, Germany, Email: pablo.delgado@audiolabs-erlangen.de
}\\
$^2$ \emph{\small Fraunhofer IIS, 90158 Erlangen, Germany,
Email: juergen.herre@iis.fraunhofer.de } } \maketitle

\thispagestyle{empty}           % weder Kopf- noch Fu�zeile auf Folgeseiten
% Beginn des eigentlichen Manuskripts
\section*{Abstract}
\label{sec:Abstract}

There is a considerable interest in developing algorithms that can predict audio quality of perceptually coded signals to avoid the cost of extensive listening tests during development time. While many established algorithms for predicting the perceived quality of signals with monaural (timbral) distortions are available (PEAQ, POLQA), predicting the quality degradation of stereo and multi-channel spatial signals is still considered a challenge. Audio quality degradation arising from spatial distortions is usually measured in terms of well known inter-aural cue distortion measures such as Inter-aural Level Difference Distortions (ILDD), Inter-aural Time Difference Distortions (ITDD) and Inter-aural Cross Correlation Distortions (IACCD). However, the extent to which their interaction influences the overall audio quality degradation in complex signals as expressed - for example - in a multiple stimuli test is not yet thoroughly studied. We propose a systematic approach that introduces controlled combinations of spatial distortions on a representative set of signals and evaluates their influence on overall perceived quality degradation by analyzing listening test scores over said signals. From this study we derive guidelines for designing meaningful distortion measures that consider inter-aural cue distortion interactions.

\section*{Introduction}
\label{sec:Introduction}

Numerous efforts have been made over the years to develop automatic perceived audio quality evaluation algorithms that incorporate models of the human hearing. These algorithms operate by comparing a reference signal (REF) against a signal under test (SUT) representing a possibly distorted version of the same signal, over which the perceived REF-SUT pair are fed into a cognitive model incorporating psychoacoustic results to produce a meaningful distortion index for a given perceptual phenomenon. Such scores are termed the Model Output Values (MOVs), following the terminology in~\cite{PEAQ}. The different MOVs representing different perceptually motivated independent distortion phenomena are then combined into a single (objective) quality score aiming to predict the value of a possible (subjective) score produced by a listening test on the same signals.

Most of these algorithms have successfully been employed for audio quality evaluation of monaural signals. Several extensions for the objective quality evaluation of binaural signals corresponding to a spatial auditory scene have been presented, but their range and effectiveness is limited due to the multidimensional nature of the spatial distortions.

Spatial audio quality can be degraded by perceived differences in localization of auditory sources, overall apparent source/scene width (ASW) or listener envelopment, among others~\cite{BlauertBook}. There is sufficient research indicating that the perceived changes in a spatial auditory scene are linked to changes in the inter-aural cues presented to the listening subject~\cite{taghipourMilan}. Therefore, many of the MOVs developed for the mentioned spatial extensions are based on measured differences of the inter-aural cues in the resulting spatial auditory scene.

With the objective of developing a meaningful cognitive model, the effect of isolated inter-aural cue distortions on overall audio quality has been investigated in basic experiments with synthetic signals and validated in application-dependent contexts such as audio coding and multi-channel audio signal processing.

%The model is tuned by extensive subjective experimentation linking a unique Model Output Value to some elicited aspect of the perceived differences in the signals reported in listening tests.

Considering the previously mentioned multidimensional aspect of the perceived spatial audio quality, a more comprehensive cognitive model should be able to additionally model how these independent inter-aural cue distortions are combined to form an overall single quality score.

The following work presents a series of experiments in order to determine the combined action of two inter-aural cue distortions, IACCD and ILDD, when the listener is faced with the task of assigning a unified overall quality score to different versions of degraded signals in comparison to a reference.

%Normally, the interaction between the different perceived spatial distortions is relegated to a regression procedure linking the different MOVs calculated over numerous reference - signal under test pairs and a unique subjective quality score for such pair. The regression model needs to be trained and validated over a sufficient amount of subjective data in order to truly reflect the relationship between MOVs and audio quality. If the amount subjective data points available is, the regression model risks overfitting. In most cases, gathering enough subjective data for the training is costly. It is therefore useful to gain a deeper understanding on the relationships of certain MOVs and audio quality before feeding the data into the regression model. This way, the extracted distortion features will represent more comprehensible models of the perception of audio quality and therefore not rely as much on the regression stage.

%MIssing.

%- We focus on auditory source width
%- What are the advantages?

\section*{Inter-aural cue modifications}

The inter-aural cue distortions are produced by modifying an implementation of the Binaural Cue Coding (BCC) perceptual audio codec~\cite{FallerBCC1}. When coding a stereo signal, the BCC encoder operates by estimating values of the signal's Inter-Channel Cross-Correlation (ICC) and Inter-Channel Level Differences (ICLD) for each frequency band of the ERB (Equivalent Rectangular Bandwidth) scale of the input. Then, the stereo input is mixed down to a single channel for data rate reduction during transmission and/or storage along the estimated inter-channel cues  as quantized meta-data. The decoder then tries to reconstruct the original auditory scene given by the original input signal from the mixdown and the estimated ICLD and ICC.

For generating the controlled distortions, the stored meta-data with the cue values is modified before the decoding process, resulting in a distorted auditory scene when compared with the decoded output without any meta-data modifications. It is assumed that distorting the inter-channel cues of the reconstructed signal will have a similar effect on the inter-aural cues. This is the case when the signals are presented through headphones because the effect of any other acoustical transfer function or cross-talk between signals and ears is considered negligible~\cite{FallerBCC1}.

The modification of the stored values is done by means of quantization/rounding with varying resolution in the quantizer steps. The human sensitivity to cue changes is not equal in all frequency regions. Therefore, the modification of the stored cue values per ERB band is carried out in a perceptually meaningful way by taking into consideration estimated sensitivity functions and Just Noticeable Differences (JNDs) of inter-aural level and inter-aural correlation cues according to the results in~\cite{doi:10.1121/1.1383296}. By combining the different cue sensitivity functions to control the quantizer resolution at each ERB band we achieve a single distortion index parameter for each cue of the ICC.

The signal modifications in ICLD and ICC were such that the degraded versions aimed to reduce the overall effect of said cues on perceived ASW and/or envelopment.

\section*{Listening Test}

A subjective test is carried out by means of a Multiple Stimulus with Hidden Reference and Anchor (MUSHRA) test~\cite{MUSHRA}. The stereo signals used for the experiment are listed in Table~\ref{tab:test_signals}. The choice of signals is inspired by previous results in overall quality perception of coding distortions~\cite{erne2001perceptual}. Perceptual audio coding algorithms tend to behave quite differently between audio signals with smooth temporal envelopes and predominant spectral harmonic structure (e.g. a violin recording), and signals with higher density of time-domain transients and flatter spectra (e.g. castanets). All recordings contain considerable reverberation to cause a clear sense of envelopment. In addition, other characteristics also perceivable in dry signals like stereo width caused by ICLD fluctuations in time are present.

\begin{table}[htbp]
    \centering
    \caption{Test signals used in the listening test}
    \vspace{2mm}
    \label{tab:test_signals}
        \begin{tabularx}{8cm}{@{\arrayrulewidth1.5pt\vline}Y@{\arrayrulewidth1.5pt\vline}Y|Y|Y@{\arrayrulewidth1.5pt\vline}}
           % \noalign{\hrule height1.5pt} \multirow{2}{*}{Variante} & \multicolumn{3}{c@{\arrayrulewidth1.5pt\vline}}{Parameter}\\
            %\cline{2-4} & A & B & C \\
            %\noalign{\hrule height1.5pt} X & AX & BX & CX\\
            \noalign{\hrule height1.5pt}
            Item ID & Description\\
            \noalign{\hrule height1.5pt}
            %\hline
            Item S1 & Reverberant solo violin recording\\
            %\hline
            Item S2 & Reverberant solo castanets recording\\
             %\hline
            Item M1 & Violin (left) and piano (right) mix\\
            %\hline
            Item M2 & Violin (left) and castanets (right) mix \\
            \noalign{\hrule height1.5pt}
        \end{tabularx}
\end{table}

Seven expert listeners in the domain of spatial audio coding were used as subjects.  The subjects were presented with the task of evaluating the overall audio quality by means of the Subjective Difference Grade (SDG) scale, using 100 points meaning no perceived impairment with respect to the reference and 0 for the highest impairment possible. The different signal treatments included combinations of three cue distortion levels introduced using the method of the previous section. For inter-channel level differences three levels are used: no ICLDD (labeled \textit{L\_null} in the subsequent analysis), mild ICLDD (\textit{L\_mid}) and highest possible ICLDD (\textit{L\_high)}, which is equivalent to not performing any ICLD reconstruction at all in the decoder. A similar procedure is carried out for generating the signal treatments related to ICCD (termed \textit{C\_null}, \textit{C\_mid} and \textit{C\_high} respectively in the next section). A mono mixdown for each of the stereo signals was used as an anchor (\textit{anchor}). In total, 9 conditions including hidden reference and anchor were presented to the subjects per test signal (i.e. item).

\section*{Results}

After the data from the listening tests was gathered, a customary post-hoc analysis was carried out. There was no need to remove any of the subjects' results in post-screening either due to low discrimination or reliability of the overall scores. The score data passed Lilliefors normality tests for all the ratings, except on ratings at the top and the bottom of the scale, where normality is not expected. The subjects reported an overall reduction of ASW and envelopment as the main effect on quality degradation, which was the intended purpose of the induced inter-channel cue distortions. However, some of the subjects also reported stereo image instability over time as a secondary cause and some timbral distortions for the highest possible ICLDD on the mix items. No other timbre distortions were reported.

A first Analysis of Variance (ANOVA) over all MUSHRA scores using fixed factors of ICLDD level, ICCD level and random factors of subject and item type rejected the null hypothesis that all conditions do not cause any significant quality degradation (i.e. the mean quality scores for all conditions are equal). No overall significant main effect of individual listener or item was confirmed. This can be explained by the fact that the full range of the scale was used for all the items (i.e. no items presented worse overall scores than others, on average). The ANOVA analysis did show significant main effects ($p < 0.05$) for ICLD and ICCD separately but no interactions among each other. However, there were significant interactions for item type-ICLDD and item type-ICCD. Based on this fact, we clustered the items to further study this interaction within different item groups. The items were divided into a first group containing solo instruments and a second group containing instrument mixes based on the similarity in their mean quality scores for each condition.

\begin{figure}[hbt]
    \begin{center}
        \includegraphics[width=9.3cm]{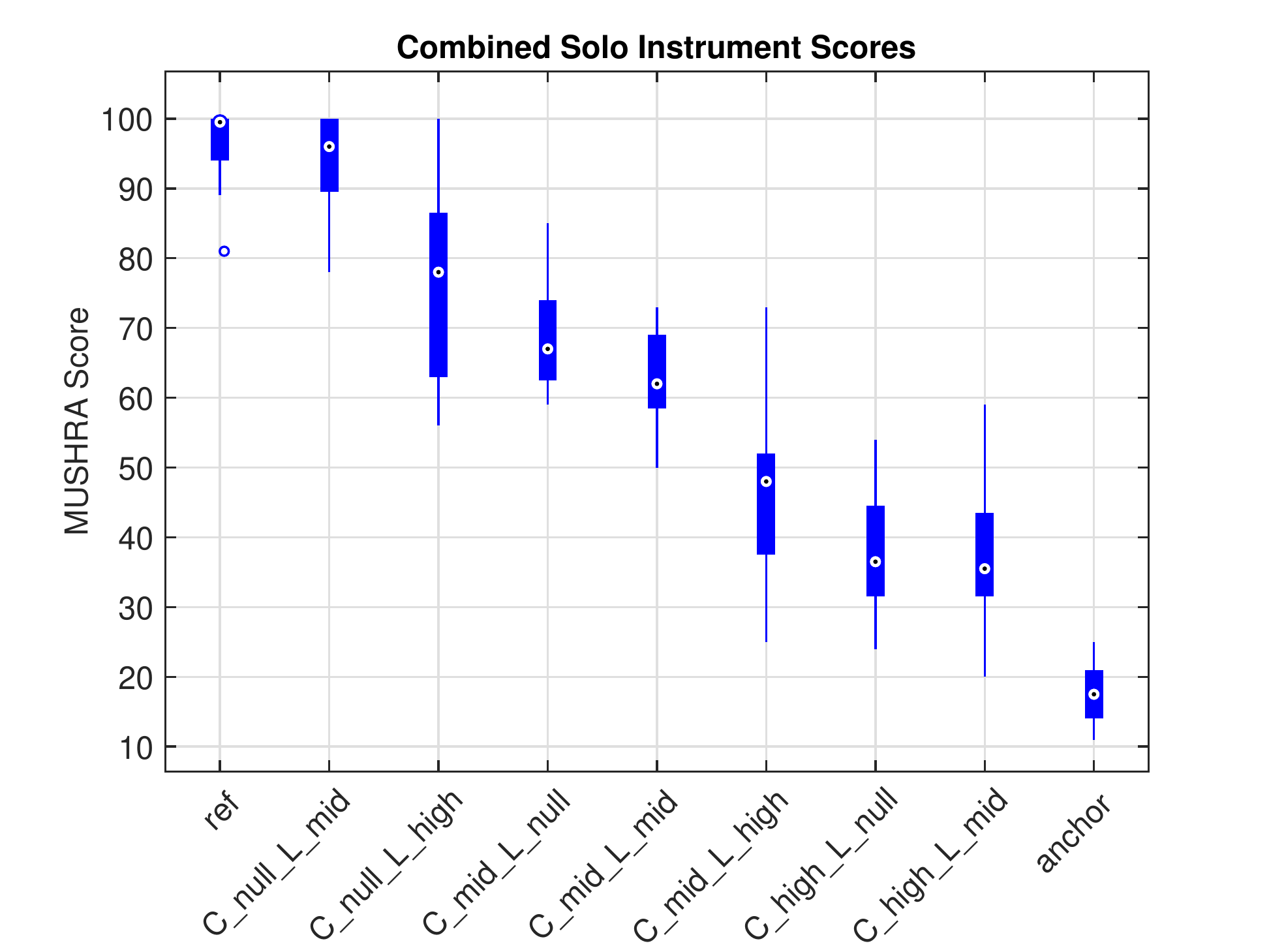}
    \end{center}
    \caption{Mean MUSHRA scores for items consisting of solo instrument recordings: Item S1 (violin) and Item S2 (castanets).}
    \label{fig:MUSHRA_solo}
\end{figure}

\begin{figure}[hbt]
    \begin{center}
        \includegraphics[width=9.3cm]{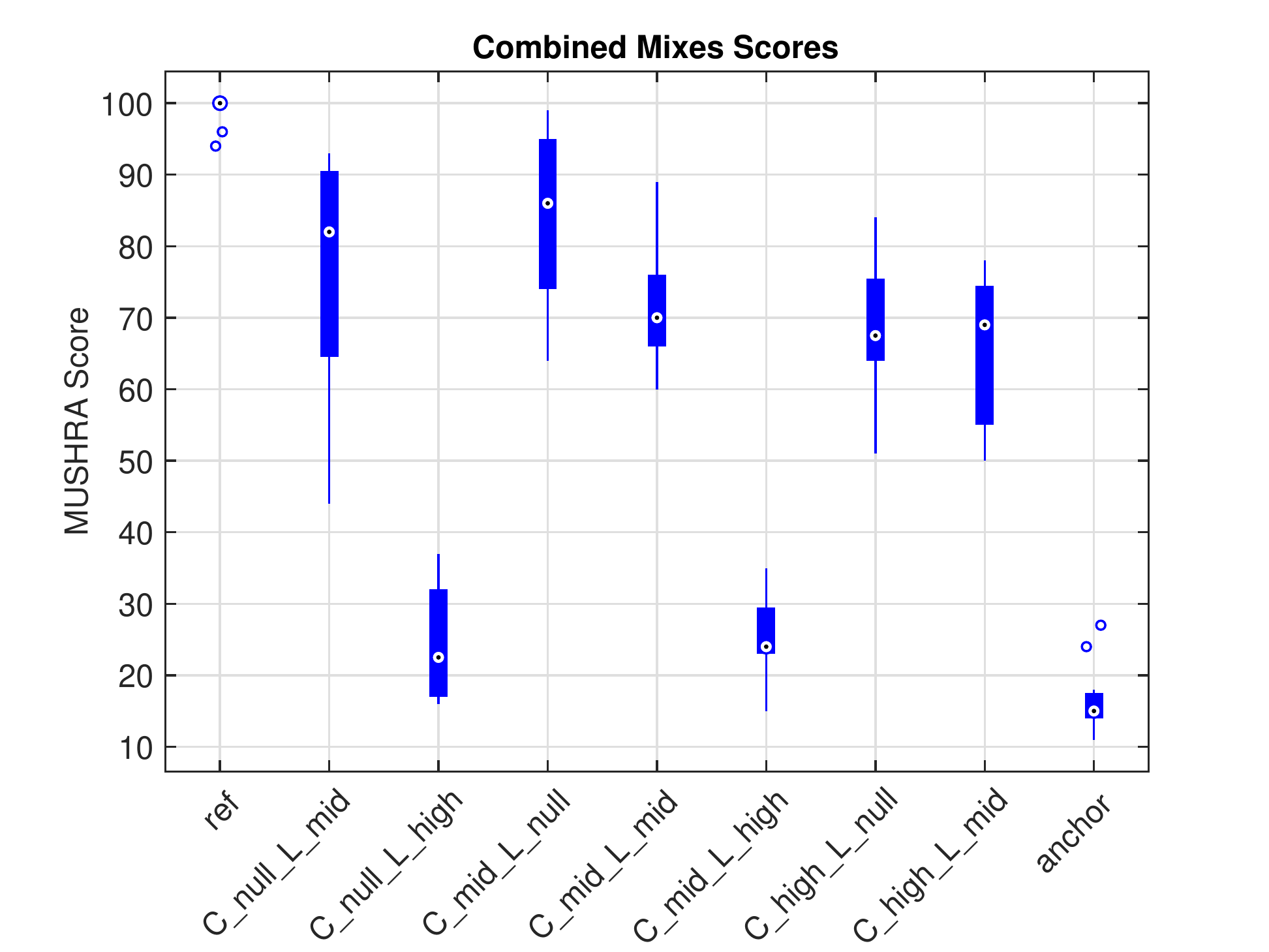}
    \end{center}
    \caption{Mean MUSHRA scores for items consisting of instrument mixes: Item M1 (violin and piano) and Item M2 (violin and castanets).}
    \label{fig:MUSHRA_mixes}
\end{figure}

Figures ~\ref{fig:MUSHRA_solo}  and ~\ref{fig:MUSHRA_mixes} show the boxplots for the combined MUSHRA scores of solo instrument items and combined scores for items with instrument mixes respectively.

Only considering the solo instrument group (Items S1 and S2), a similar ANOVA analysis showed significant ICCD and ICLD main effects, but no interaction between the factors. Post-hoc pairwise t-tests with the Bonferroni correction showed significant differences between \textit{L\_high} and \textit{L\_null--L\_mid} ICLDD for all ICCD conditions and no significant interactions for all three ICCD levels.

Considering the instrument mixes group (Items M1 and M2), there were again significant main effects of ICCD and ICLD, and significant interactions between the two. Pairwise t-tests showed significant differences between \textit{L\_high} and \textit{L\_null--L\_mid} ICLDD for all ICCD conditions and only significant mean differences for \textit{L\_null} and \textit{L\_mid} in the case of no ICCD.

\section*{Discussion}

\begin{figure}[hbt]
    \begin{center}
        \includegraphics[width=8.7cm]{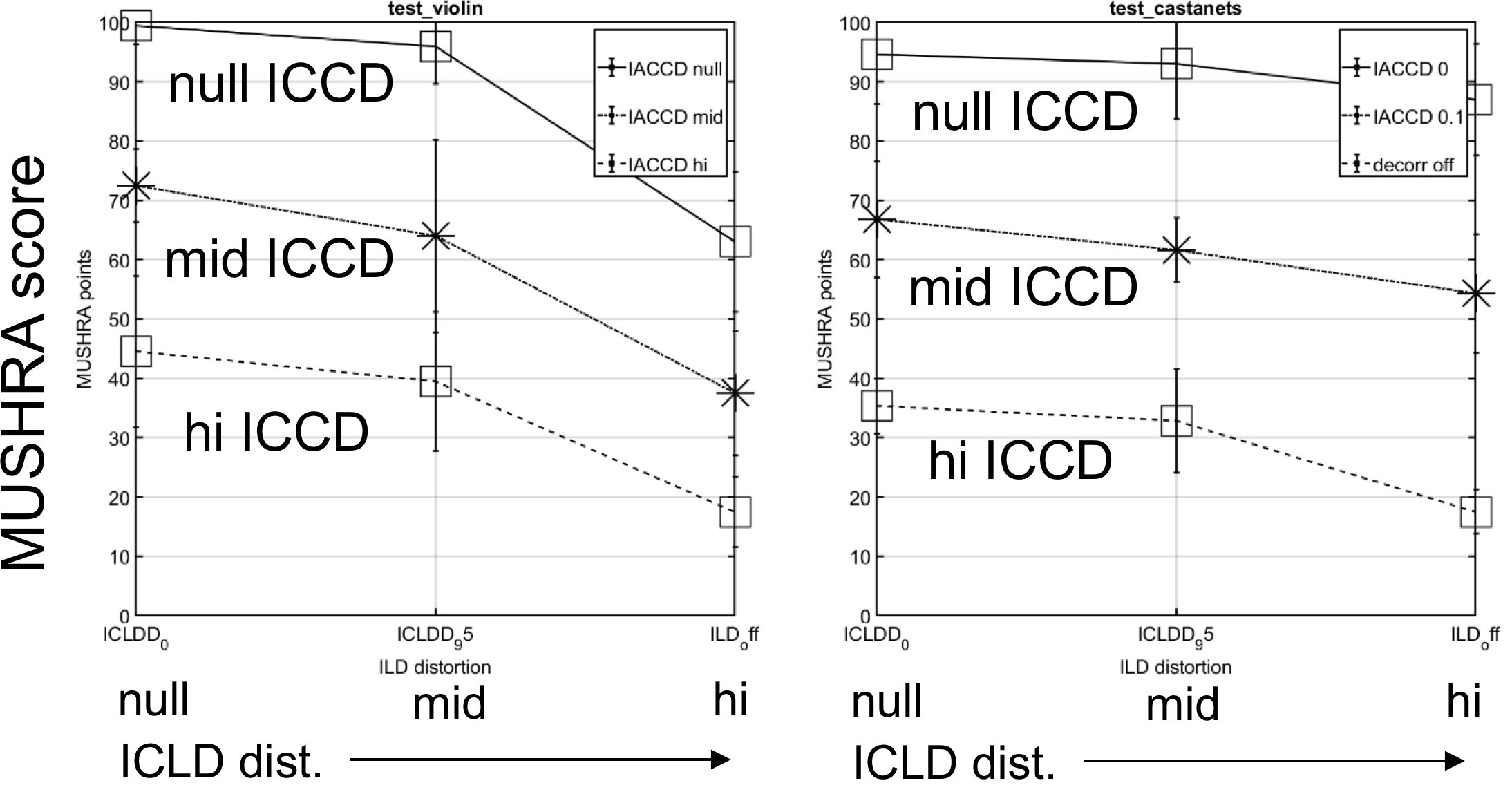}
    \end{center}
    \caption{Pooled mean MUSHRA scores for solo instruments according to different ICLDD levels into three different ICCD level curves with 95\% confidence intervals. Left plot: Item S1 (violin). Right plot: Item S2 (castanets).}
    \label{fig:Viol_Cast_Interact}
\end{figure}

Pooling the treatments (conditions) according to increasing ICLDD in three different ICCD level curves for the items in the first group (Figure ~\ref{fig:Viol_Cast_Interact}) reveals a tendency to higher ICLDD sensitivity in the violin recording (S1) than in the castanets recording (S2). Additionally, there is a marginal higher degradation of quality due to ICCD in the castanets recording than in the violin recording. A visual inspection hints to the fact that the type of signal determines the predominance of one cue distortion with respect to the other for this particular group. However, these differences cannot be confirmed from the ANOVA analysis or pairwise t-tests (no interactions reported) and further studies with more diverse data are needed.

\begin{figure}[hbt]
    \begin{center}
        \includegraphics[width=8.7cm]{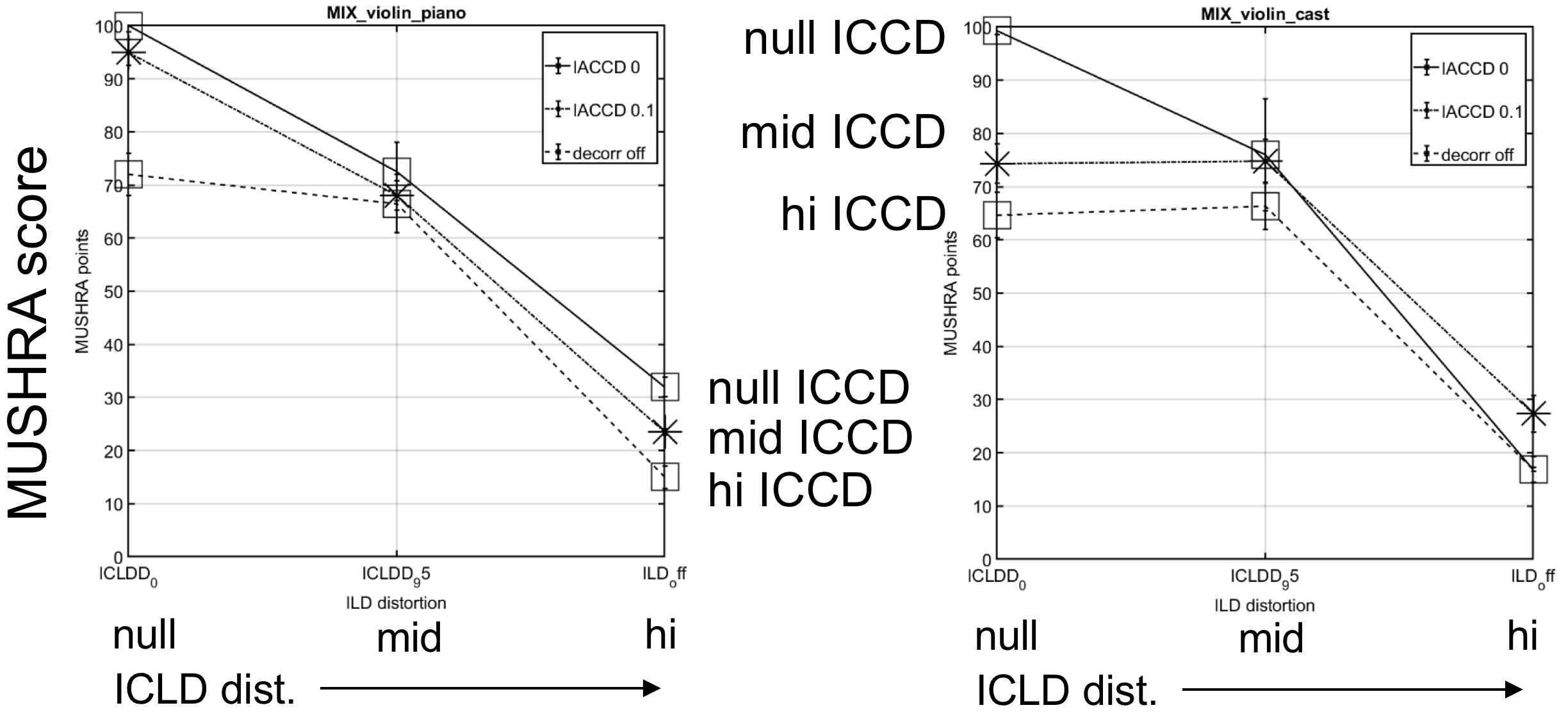}
    \end{center}
    \caption{Pooled mean MUSHRA scores for instrument mixes according to different ICLD levels into three different ICCDD level curves with 95\% confidence intervals. Left plot: Item M1 (violin and piano). Right plot: Item M2 (violin and castanets).}
    \label{fig:Mix_Interact}
\end{figure}

The same pooling can be seen in Figure ~\ref{fig:Mix_Interact} for the second group containing instrument mixes. In this case, the ICLDD influence on overall quality has more impact than in the case of the first group. In comparison, the ICCD influence is not as strong; all three ICCD curves are very close to each other with the exception of the case in which no ICLDD is present at all. In this case, inter-channel correlation changes in the signal represented by ICCD do make a significant difference in quality. As soon as there is some ICLDD, the effect of ICCD becomes weaker. This is confirmed by the pairwise t-tests reported in the previous section.

In order to test the remarks from the visual inspection we can perform an effect strength analysis in terms of Hedges' $g$ coefficient between the mean reference condition scores (no distortion present) and the mean scores of all distorted items. There is indeed a stronger effect size of ICCD for the solo instruments ($g=2.73$)  than for the instrument mixes ($g=1.77$). As for ICLDD, they have a stronger effect size on overall quality for instrument mixes ($g=2.11$) than for solo instruments ($g=1.36$). Considering effects on each group, ICLDD and ICCD have comparable effect sizes in the case of solo instruments, whereas ICLDD effect sizes are considerably higher than ICCD effect sizes on instrument mixes.

In solo instrument recordings, the simple auditory scene composed by one object with prominent room reverberation allows the listeners on one hand to increase focus on variations in the perceived envelopment and ASW differences in equal amounts. On the other hand, two or more distinct objects at very different perceived directions might take the listeners' attention away of --or even mask-- the effects of distortions in measurements associated with perceived room changes or envelopment.

The different distortion effect strengths on the instrument mixes' quality in comparison to solo instruments might hint to a relative importance of the two perceived attributes (ASW and envelopment) in the overall audio scene. The research field known as Auditory Scene Analysis (ASA) proposes that the interpretation of the individual binaural cue behaviors needs to be considered in the context of the observed auditory input signal characteristics ~\cite{WangASA}. This is especially important with increasing scene complexity, where superimposed auditory objects and their interaction with noise and reverberation play a larger role. It is expected that predicting spatial audio coding quality contemplates auditory scenes within a wide range of complexity levels. Following this line of thought, the techniques developed in ASA could be advantageous in the development objective psychoacoustic measurement models.

\section*{Conclusions}

The influence of two inter-channel cue distortions, inter-channel level difference distortions (ICLDD) and inter-channel correlation distortions (ICCD) on overall audio signal quality degradation was studied. A series of MUSHRA listening tests on signals of varying characteristics presented through headphones was carried out along with a subsequent statistical analysis including effect sizes of both types of distortions and their combinations.

Although the diversity of the tested signals for this first study was limited, results showed that these can at least be divided into two groups according to ICLDD and ICCD effect on overall quality: solo instruments and instrument mixes. For solo instruments -- which contain one single auditory object -- effect sizes of ICLDD and ICCD are about the same. Moreover, the effects of ICLDD and ICCD are independent of one another. Room reverberation is prominent on these solo instrument recordings, and distortions in inter-channel correlation might be easily perceived as changes in listener envelopment.

In the case of instrumental mixes containing at least two auditory objects at the same time, the effects of ICLDD on overall quality degradation are significantly larger than ICCD effects. Distortions introduced in the mix signals are also dependent on one another. For example, ICCD has a significantly higher influence on overall quality as long as no ICLDD is present. Once, ICLDD is present, ICCD causes no significant quality degradation. The main effects and interactions of the inter-channel cue distortions might further change as the auditory scene increases in complexity. This case remains for a later study.

Considering the development of perceptual models of the human hearing for the task of objective quality estimation of spatial audio, the signal-dependent nature of inter-channel cue distortions suggests the use of higer-level inter-aural distortion analysis stages that consider the nature of the binaural input signals for determining the distortions' effect on quality. These higher-level distortion analysis modules might incorporate concepts of Auditory Scene Analysis or binaural cue combination mechanisms as extensions of a cognitive model.

%\subsection*{Literaturverzeichnis}
%Verwendete Literatur wird am Ende des Manuskripts angegeben. Artikel %~\cite{ArticleReference}, B�cher ~\cite{BookReference} und Internet-Adressen %~\cite{URLReference}~\cite{PDFCreator}~\cite{Ghostware} werden wie unter Literatur %angegeben zitiert.
%\subsection*{Formeln}
%Gleichungen sind zu nummerieren und so anzuordnen, wie beispielsweise %Gleichung~(\ref{eq:FirstEquation}).
%\begin{flalign}
%\label{eq:FirstEquation}
%           \qquad\qquad F &= \pi r^2 & \mathrm{\left[m^2\right]} \quad
%\end{flalign}
%Stellen Sie sicher, dass alle vorkommenden Variablen bei der ersten Verwendung erl�utert werden.\\
%Benutzen Sie f�r Gleichungen statt der equation-Umgebung die flalign-Umgebung aus dem amsmath-%Package.

\input{where_is_the_bibfile.tex}
\bibliographystyle{abbrv}
%\bibliography{../../../papers_st_polqa}

%\begin{thebibliography}{5}

%\bibitem{ArticleReference}
%Schall, A.: How to write a manuscript. Acta Acustica united with
%Acustica 90 (2004), 2203-2503
%\bibitem{BookReference}
%Klang, B.: Akustik im �berblick. Schall und Rauch Verlag, Stadt,
%2010
%\bibitem{URLReference}
%DAGA 2019 Homepage, URL:\\
%\url{www.daga2019.de}
%\bibitem{PDFCreator}
%PDFCreator, URL:\\
%\url{http://sourceforge.net/projects/pdfcreator}
%\bibitem{Ghostware}
%Free software Ghostview and Ghostscript, URL:\\
%\url{http://www.cs.wisc.edu/~ghost/}

%\end{thebibliography}

\end{document}

%% file: where_is_the_bibfile.tex
\bibliography{./papers_st_polqa}